\documentclass{ametsocV6.1}

\title{Do Climate Models Need Microphysical and Convective Parameterizations to Generate Accurate Precipitation Fields?}

\authors{Raul Moreno\aff{a}, Dale Durran\aff{a}\correspondingauthor{Raul Moreno, morenor@uw.com}}

\affiliation{\aff{a}{University of Washington}}
\def\mle{ML$_{\rm ERA5}$\,}
\def\mli{ML$_{\rm IMERG}$\,}
\abstract{Accurately representing surface precipitation is crucial for the operational use of weather and climate models. Presently, global numerical weather prediction (NWP) models struggle to accurately generate precipitation due to their parametrization of unresolved deep convective clouds and, in regions of grid-resolved ascent, inadequate parameterizations of cloud microphysics. Here we 
bypass these parameterizations with a machine learning model that diagnoses precipitation from 13 ERA5 fields that are easily observed and assimilated, as opposed for example, to fields like rain or cloud liquid water. We train a pair of models; \mle using ERA5 precipitation as the target, and \mli using a satellite based precipitation product. \mle closely reproduces the ERA5 precipitation at all intensities. When evaluated against the satellite dataset, \mli closely matches observations, notably reproducing the diurnal cycle of the satellite product. \mli generally captures extremes better than ERA5 while also reducing ERA5's  overproduction of light precipitation. When evaluated against a third ground-and-radar-based dataset, \mli inherits the strengths of the satellite dataset which is superior to ERA5 in the summer months.} 

\begin{document}

\maketitle

%
%
%
%
%
%

%




\section{Introduction}
Precipitation is one of the most important fields in weather and climate modeling. Reliable predictions of floods and droughts, snow and ice storms, and long-term hydrological trends are essential for anticipating and mitigating weather and climate impacts. Precipitation is, however, one of the most difficult fields to capture in global numerical weather prediction (NWP) models because of its high intermittency in space and time and the complex multiscale dynamics governing its formation, much of which occurs at the subgrid scale. Even high resolution km-scale models rely on synthetic microphysical variables to approximate precipitation processes. This results in systematic biases and unrealistic representations that degrade both short-term weather forecasts and climate simulations.

\begin{figure}[h!]
\begin{center}
    \includegraphics[width=0.8\textwidth]{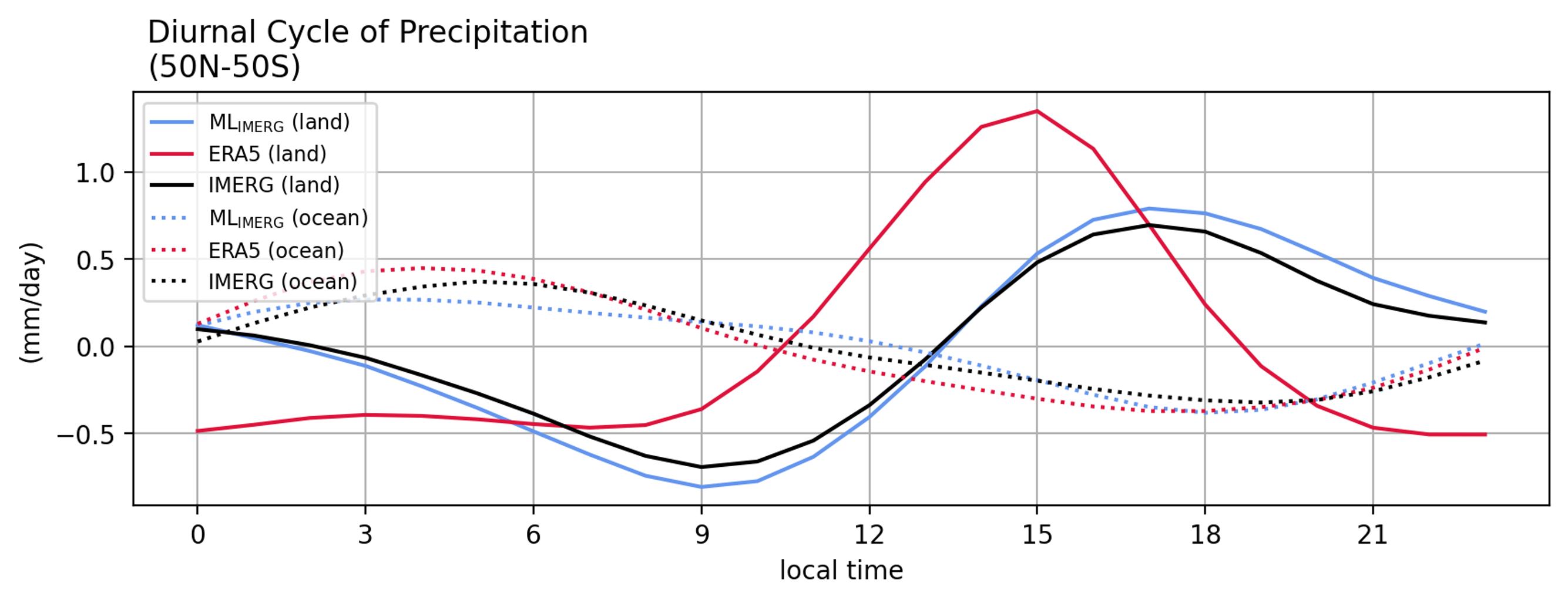}
\end{center}
\caption{Diurnal cycle of precipitation rate anomalies relative to the local daily mean values for 2022-2024 over 50N-50S for ERA5 (red), IMERG (black) and our ML model trained on the IMERG data (blue). Solid lines are precipitation over land and dotted is precipitation over the ocean.}
\label{intro_diurnal}
\end{figure}

Traditional global NWP models like the European Center for Medium Range Weather Forecasting’s (ECMWF) Integrated Forecasting System (IFS) generate precipitation through two parameterizations, one representing sub-grid-scale deep convection and the other microphysical processes occurring in response to adiabatic cooling during grid-resolved vertical motions \citep{european_centre_for_medium-range_weather_forecasts_ifs_2024}.  These microphysical processes, governing the formation and dissipation of clouds and precipitation, are largely represented by 
substantial simplifications of more complex hydrometeor distributions. For example, the continuous spectrum of liquid water droplets is simplified into just two bins ``cloud liquid specific water content” CLWC and ``precipitation rain specific water content” CRWC \citep{european_centre_for_medium-range_weather_forecasts_ifs_2024}. The convective parameterization estimates vertical transport of heat and moisture at the subgrid level. In the IFS this is implemented with a mass-flux scheme triggered in unstable conditions and tuned using empirically derived estimates of entrainment and detrainment \citep{european_centre_for_medium-range_weather_forecasts_ifs_2024}. One defect of the convective parametrization is that it misrepresents the timing of convection, with weaker convection occurring too early and resulting in too much light precipitation \citep{becker_characteristics_2021, stephens_dreary_2010, demott_convective_2007}.
The approximations used to parameterize these processes, combined with the lack of assimilated observations in the total precipitation field yield known biases in the IFS forecasts as well as in the popular reanalysis dataset ERA5, which uses the IFS as its base model. Timing errors in the daily cycle of precipitation in ERA5 are shown in comparison to satellite precipitation measurements from (IMERG) in Fig. \ref{intro_diurnal}.

In addition to timing errors in the diurnal cycle, the IFS shows an overall wet bias in forecasts when evaluated against observations \citep{lavers_precipitation_2021}. A similar result is observed in the ERA5 dataset which also shows difficulty capturing extreme precipitation events \citep{lavers_evaluation_2022}. The largest errors in ERA5 are found in the tropics, moving seasonally with the intertropical convergence zone \citep{lavers_evaluation_2022}. The wet bias in the IFS has been attributed to an overproduction of light precipitation and it has been suggested that in non-cloud-resolving models this is the result of inadequate convection parameterizations, which are characterized by an early energizing of the boundary layer and a premature triggering of convection \citep{demott_convective_2007, stephens_dreary_2010}.
These issues are not completely eliminated in finer-scale models that resolve convection. For example, analysis of precipitation from a set of convection-permitting models in the DYAMOND project revealed that even when convective storm systems are directly simulated, convective rainfall intensity was overestimated while stratiform rainfall was underestimated, potentially due to inadequate microphysics parameterizations and representation of entrainment \citep{feng_mesoscale_2023}.

In the last decade, the application of deep learning to weather prediction has undergone rapid development, with many purely data-driven models achieving or even surpassing the skill of NWP in measures like RMSE and ACC at short to medium range predictions \citep{bi_accurate_2023, lam_learning_2023, lang_aifs_2024,Moldovan2025AIFSupdate}. 
Some models have achieved indefinite stability, demonstrating competency in a wider range of applications such as modeling the current climate \citep{watt-meyer_ace_2023, Cresswell-Clay2025DLESyM}. Training these models requires large amounts of data such as the widely used ERA5 reanalysis. While accurate for easily observed and assimilated fields such as 500-hPa height, the ERA5 precipitation data is less optimal because it is obtained from IFS forecasts without directly assimilated observations of precipitation. Using such model-derived data as the training target will propagate biases to the data-driven model.

Recently, deep learning models have been developed that estimate precipitation directly from observations. High-resolution models like MetNet-2 and Global MetNet trained on satellite observations outperform state-of-the-art HREF ensemble predictions at lead times up to 12 hours \citep{Espeholt2022DeepLearning12hPrecip,Agrawal2025OperationalDeepLearningSystemGlobalNowcasting}. Precipitation observations have also been used to train deep learning models for correction-related tasks such as postprocessing of forecasts as well as ML-driven data assimilation \citep{hess_deep_2022,Xu2025FuXiDA}. Satellite observations were used to train the precipitation module in the dynamical core-ML hybrid model NeuralGCM, outperforming the ECMWF ensemble at medium-range forecasting \citep{yuvalNeuralGCM_precip2026}. Generative models such as DiffObs have also shown realistic estimates of precipitation by training on satellite observations \citep{stock_diffobs_2024}. 

Rather than using machine learning to directly forecast precipitation, we focus on the fundamental question of whether the deep convection and microphysical parameterizations used in the IFS to create the ERA5 precipitation analysis are unnecessary in the sense that other more easily assimilated ERA5 variables carry all the information necessary to reconstruct the precipitation field.  As such, we first endeavor to compute precipitation fields essentially equivalent to those in ERA5 by a deterministic machine learning model using thirteen 2D ERA5 fields that are easily obtained through observation and data assimilation on the synoptic scale. Only three of these fields are directly influenced by atmospheric humidity: total column water vapor, out going top-of-atmosphere thermal radiation, and specific humidity at 850 hPa. Our focus will be on resolutions in space and time appropriate for typical climate model simulations: a $1^{\circ}\times 1^{\circ}$ latitude-longitude global mesh and 3- or 24-h accumulations.  Our approach can be contrasted with that in \cite{Gettelman2021}, where a machine learning model was developed to emulate details of the warm rain process in the NCAR's Community Atmospheric Model, CAM6 on a 0.9$^{\circ}\times1.25^{\circ}$ latitude-longitude grid.  Instead of learning a fast representation of a stochastic bin model of warm-rain precipitation process, we explore whether the large-scale atmospheric fields themselves are sufficient to determine the total precipitation

After examining the extent to which ERA5 fields can be used to determine ERA5 precipitation without input from microphysical or deep convective parameterizations, our second goal is to train and analyze a similar precipitation model using NASA’s Integrated Multi-satellitE Retrievals for GPM (IMERG). Training on observational data provides an avenue for improving precipitation forecasts without relying on advancements to microphysical or convective parameterizations. As shown in Fig. \ref{intro_diurnal}, this approach can be used to remove the bias in the diurnal cycle of precipitation from the ERA5 data. As a further benchmark, we will also assess the performance of this model against a second observationally based precipitation dataset, the National
Centers for Environmental Prediction (NCEP) Stage IV data \citep{LinMitchell2005}.

In previous work, \citet{Larraondo2019} presented machine learning models that estimated precipitation in the ERA-Interim dataset over Europe from geopotential height at roughly 80-km resolution.  While it yielded promising results, their verification of the precipitation was very basic.  In contrast to \citet{Larraondo2019}, we consider a more complete set of metrics used in typical precipitation assessments over the full globe. 
\citet{Ji2024LeadseePrecip} used machine learning to estimate global NOAA CMORPH precipitation data {\citep{Joyce2004_CMORPH}} upscaled to 0.25$^{\circ}$ resolution from 69 ERA5 fields. Here we use roughly an order of magnitude fewer inputs, and consider the extent to which ERA5's own precipitation field can be determined by those inputs.  We also verify models trained to match precipitation from one observed dataset (IMERG) against a second observed dataset, NCEP's Stage IV precipitation analysis.

The remainder of this paper is organized as follows. In section 2 we describe our model, data sources, and briefly review several performance metrics particularly appropriate for assessing precipitation. Section 3 analyzes the performance of a model trained to forecast ERA5 precipitation from 10 ERA5 fields. Section 4 evaluates the performance of a similar model trained against IMERG precipitation data. Section 5 presents our conclusions.

\section{Methods and Data}
\subsection{Model Training}

\begin{figure}[h!]
\begin{center}
    \includegraphics[width=\textwidth]{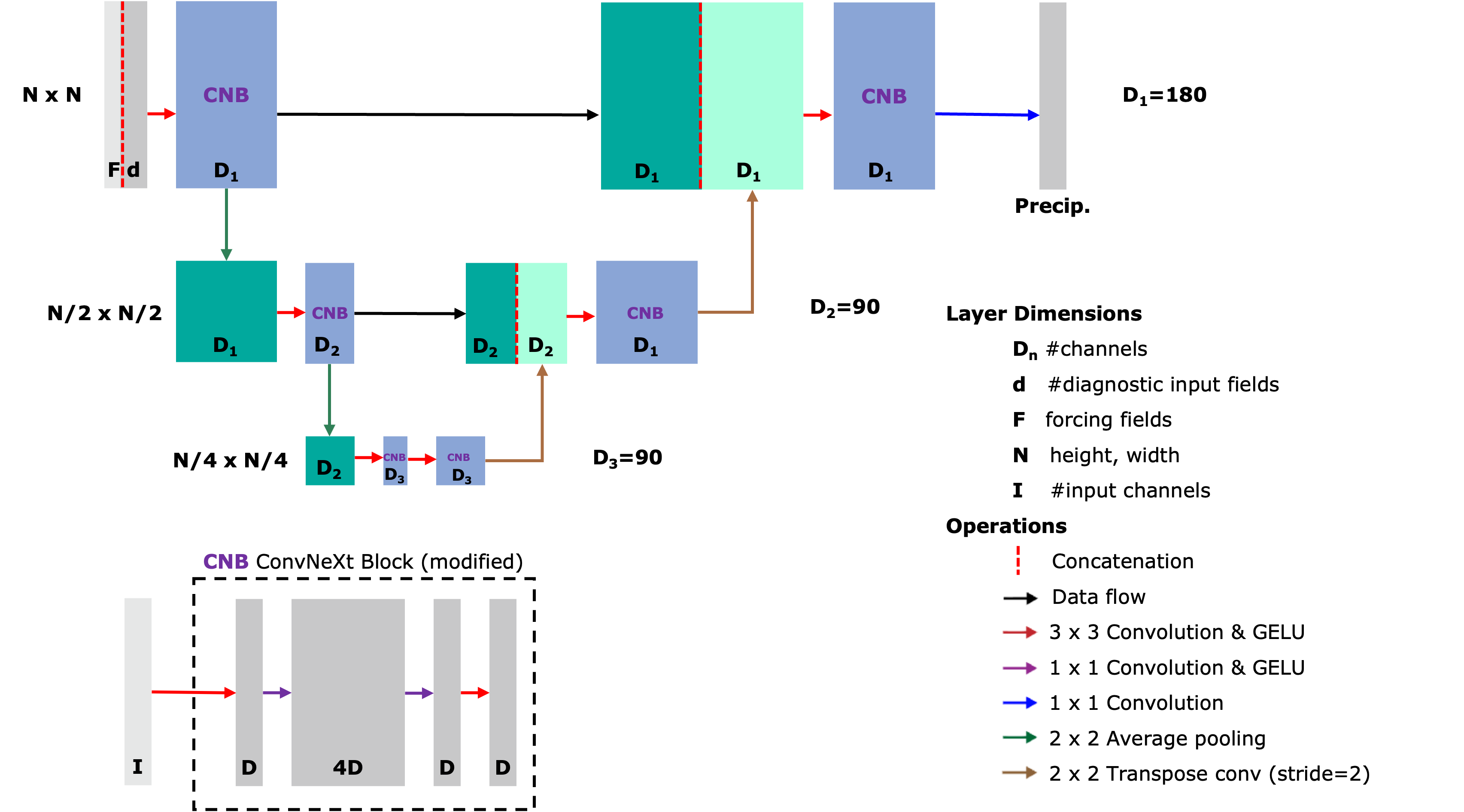}
\end{center}
\caption{Diagram of CNN U-net used for the ML precipitation model. ConvNeXt block (blue block) layout shown in bottom left. Dark green blocks represent layers evaluated by the encoder while the lighter green blocks represent layers evaluated by the decoder.}
\label{unet}
\end{figure}

The ML-based precipitation model presented here uses a U-net style convolutional neural network (CNN) which is a slightly modified version of the U-net used for the Deep Learning Earth Systems Model (DL\textit{ESy}M) by \citet{Cresswell-Clay2025DLESyM}. The U-net (Fig.~\ref{unet}) consists of 3 levels and uses ConvNeXt blocks which expand the latent layer depth by a factor of four using a $1\times1$ spatial convolution. The U-Net reverses the typical filter depth distribution, allocating more filters to higher-resolution features before downsampling in the deepest two layers. \citet{karlbauer_advancing_2024} notes that this arrangement is practical for ML weather models as it allows more capacity to be spent on capturing fined grained features---this is especially relevant for the precipitation variable. Here we use a channel depth arrangement of 180-90-90. Data is input for the two times bounding the 3-hour period over which the accumulated precipitation is diagnosed. 

An important consideration for training machine learning models with precipitation data is its significant right-skewness because cases of no precipitation are far more common than cases with precipitation. If trained on the raw data, the model tends to predict only zero or very small precipitation totals \citep{rasp_datadriven_2021}. We therefore log-transform the precipitation variable in a preprocessing step before training, such that 
\begin{align*}
    \tilde{P} = \ln(1 +P/\epsilon), 
\end{align*}
where $P$ is precipitation in m, and $\epsilon = 1 \times10^{-8}$ m.  This  transformation was pivotal to successfully training our model.

Our MSE training loss is weighted to emphasize precipitation extremes.
\begin{align*}
    \text{Loss} = \frac{1}{N} \sum_{(x, y)} w* (\tilde{P} - \tilde{P}_{\text{obs}})^2 \quad \text{if } \tilde{P}_{\text{obs}} > 0 ,
\end{align*}
where 
\begin{align*}
    w=
    \begin{cases}
        e^{b \tilde{P}_{\text{obs}}}, & \text{if } \tilde{P}_{\text{obs}} > 0 \\ 
0.15, & \text{if } \tilde{P}_{\text{obs}} = 0
    \end{cases}
\end{align*}
Non-precipitating events are down weighted, while heavier events receive additional exponential weighting with $b=0.16$. This further emphasizes the extreme end of the precipitation distribution and proved crucial to achieving skill at the highest thresholds.

\subsection{Data}
A complete list of input and target variables appears in Table \ref{tab:variables}. The input variables were initially chosen to match the DL\textit{ESy}M model output variables \citep{Cresswell-Clay2025DLESyM}. Three additional variables (horizontal winds and specific humidity at 850 hPa) were added to better characterize the water vapor transport.  Nevertheless, the model can be applied 
To NWP forecasts or the ERA5 data itself, and to examine the maximum potential of our model we will diagnose precipitation directly from ERA5 inputs rather than a forecast. The target variables consist of precipitation from ERA5 or from IMERG. For the model trained on ERA5, the training set spans 1980-2018, 2019-2021 for validation, and 2022-2024 for testing. For the model trained on IMERG, the training set spans 2000-2018, with the same validation (2019-2021) and testing (2022-2024) periods.

\begin{table}[t]
\begin{center}
\begin{tabular}{p{7cm}p{1.3cm}p{4cm}p{1.8cm}}
\topline

\textbf{Variable Name} & \textbf{Source} & \textbf{Preprocessing} & \textbf{Type} \\
\midline
1000-hPa geopotential height (z1000) & ERA5 & & Input \\
500-hPa geopotential height (z500) & ERA5 & & Input \\
250-hPa geopotential height (z250) & ERA5 & & Input \\
700-300 thickness ($\tau_{300-700}$) & ERA5 & $Z_{300} - Z_{700}$ & Input \\
2-m temperature (T2m) & ERA5 & & Input \\
850-hPa temperature (T850) & ERA5 & & Input \\
850-hPa specific humidity (q850) & ERA5 & & Input \\
850-hPa meridional winds (U850) & ERA5 & & Input \\
850-hPa zonal winds (V850) & ERA5 & & Input \\
Total column water vapor (TCWV) & ERA5 & & Input \\
10-meter wind speed (WS10) & ERA5 & & Input \\
Top of atmos. thermal radiation (TTR) & ERA5 & Divided by time. & Input \\
Sea surface temperature (SST) & ERA5 & & Input \\
3-hr (model) total precip. (TP3) & ERA5 & 1-hr summed to 3-hr & Output \\
3-hr (satellite) total precip. (TP3) & IMERG & 1-hr (30-min samples) summed to 3-hr & Output \\
\botline
\end{tabular}
\end{center}
\caption{List of variables, their sources, preprocessing, and whether they are model inputs or outputs.}
\label{tab:variables}
\end{table}

\subsubsection{ERA5}
We use ECMWF Reanalysis version 5 (ERA5) at 3-hourly resolution for our model inputs \citep{hersbach_era5_2020}. Additionally, topography and a land-sea mask are supplied as constant fields. To obtain a 3-hour precipitation accumulation field, the 1-hour field is aggregated so that each sample represents the total precipitation accumulated over the 3 hours ending at the time of the sample. This field is also at 3-hour resolution.
\subsubsection{IMERG}
Satellite observations are from NASA’s Integrated Multi-satellitE Retrievals for GPM (IMERG) version 7 dataset, which includes precipitation fields every 30 minutes from 2000 to present. IMERG combines microwave, infrared, and rain gauge estimates of precipitation to provide a `best estimate' record of precipitation at the surface \citep{huffman_nasa_2020}. Like the ERA5 precipitation data, at 3-hour intervals the IMERG 1-hour accumulation data (estimated from the hourly rates at half-hourly samples) was aggregated to 3-hour accumulations. 
\subsubsection{Stage IV}
A ground-based observational dataset was also used for additional validation. The National Centers for Environmental Prediction (NCEP) Stage IV data provides radar and rain-gauge estimates of precipitation \citep{LinMitchell2005} and has previously been used as a reference dataset for validation of other precipitation estimates \citep{cui_can_2020}.
\subsection{Evaluation}
Skill metrics for precipitation require special attention; RMSE, for example, is susceptible to ``double penalty” errors, which excessively penalize a forecast for slight displacements of an otherwise good representation of the true precipitation distribution \citep{necker_fractions_2024}. To minimize this, we evaluate the fraction skill score (FSS), which compares the proportion of grid points within a defined area where precipitation surpasses a given threshold to the corresponding proportion in the verifying data \citep{roberts_scale-selective_2008}. For this paper, all FSS evaluation is computed using a neighborhood size of $5\times5$ points for each of six different intensity bins. In the FSS, a score of one indicates a perfect forecast, while a score of zero is that of the worst-case forecast.

The Stable Equitable Error in Probability Space (SEEPS) score is a climatologically considerate metric. The SEEPS score measures error in probability space and assesses estimates of dry, light, and heavy precipitation, as  defined by climatology at each location \citep{Rodwell2010}.  We plot 1-SEEPS, which gives a score of one for a perfect forecast and an expected value of zero for an unskillful forecast \citep{northAssessmentSEEPSSEDI2013}.

We also evaluate the Equitable Threat Score (ETS) as a function of precipitation intensity, where
\begin{equation}
    {\rm ETS}=\frac{{\rm hits}-E}{{\rm hits}+{\rm misses}+{\rm false\; alarms}-E}
\end{equation}
and $E$ is the expected number of hits over the same number of $n$ forecasts generated by a random forecast system with the same average rate of occurrence as the actual forecasting system
\begin{equation*}
    E=({\rm hits}+{\rm false\; alarms})\times ({\rm hits}+{\rm misses})/n
\end{equation*}

We evaluate the presence or absence of precipitation above specified threshold values to compute the probability of detection (POD= hits/[hits+misses]) and the success ratio (SR=hits/[hits+false alarms]) at every individual point over the full 2022-2024 test set. These point-wise values are then spatially averaged and POD is plotted as a function of SR for each threshold, generating a performance diagram \citep{Roebber2009}, for which excellent scores are plotted in the upper right near (1,1).
We round out our verification metrics with spatial maps of the temporal correlation between the forecasts and the verifying data over the 2022-2024 test period. 

\section{Results: Reproducing ERA5 Precipitation from ERA5 Inputs}
\label{sec:results}
In this section, we analyze the global performance of a model (\mle) trained to reproduce ERA5 3-hr precipitation from the 13 input fields in Table~\ref{tab:variables}. This allows us to assess the degree to which the ERA5 precipitation fields can be determined by a small collection of easy-to-assimilate fields in contrast to a more conventional combination of microphysical and deep convective parameterizations.

\begin{figure}[h!]
\begin{center}
    \includegraphics[width=\textwidth]{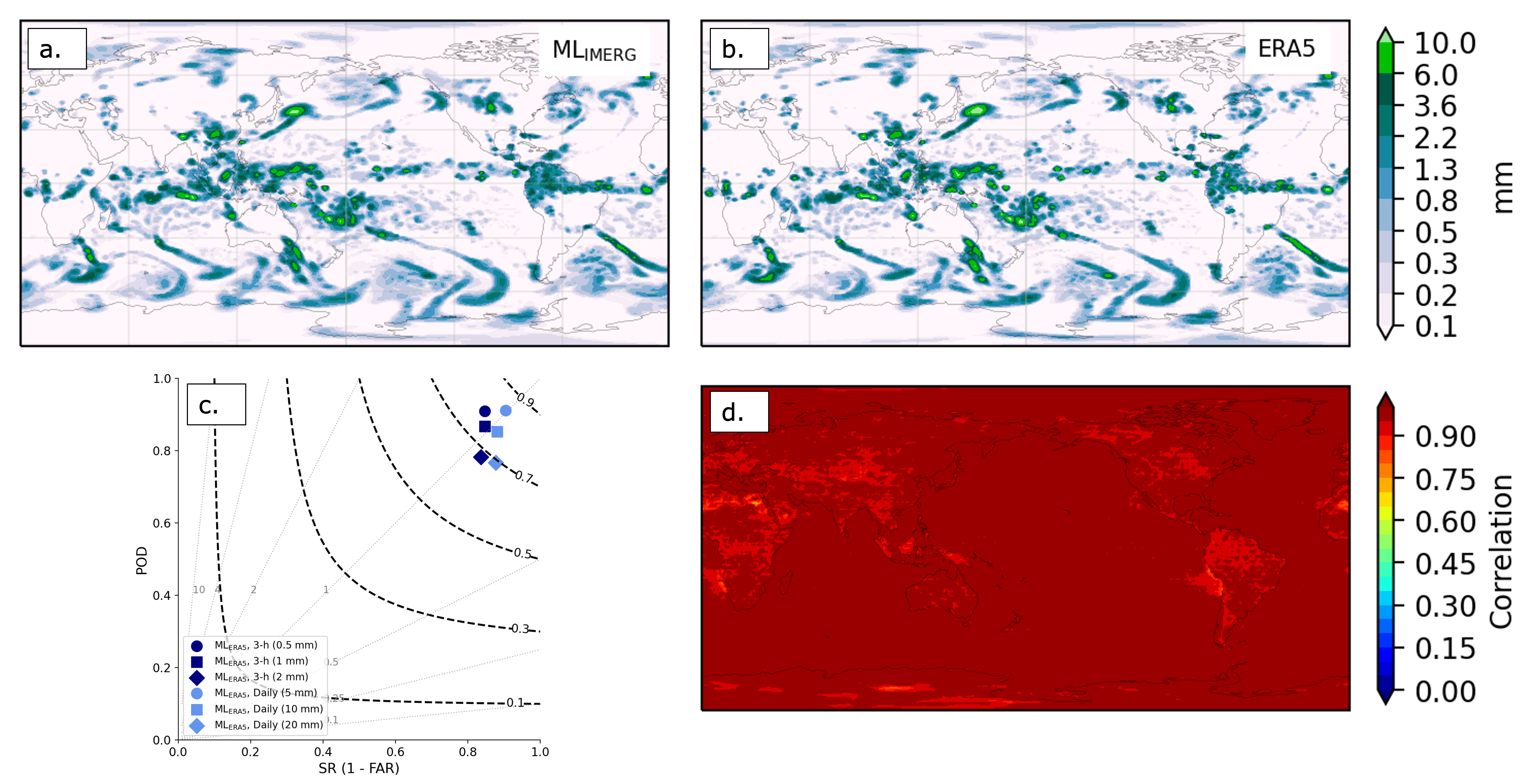}
\end{center}
\caption{Example 3-hour totals beginning at 00 UTC on at April 30, 2022 for (a) \mle and (b) ERA5. (c) Performance diagram for \mle 3-h totals (navy) and daily totals (light blue) showing skill by intensity threshold. (d) Temporal correlation of 3-h totals from \mle with ERA5 estimates.}
\label{era5-era5_ex}
\end{figure}

\begin{figure}[h!]
\begin{center}
    \includegraphics[width=\textwidth]{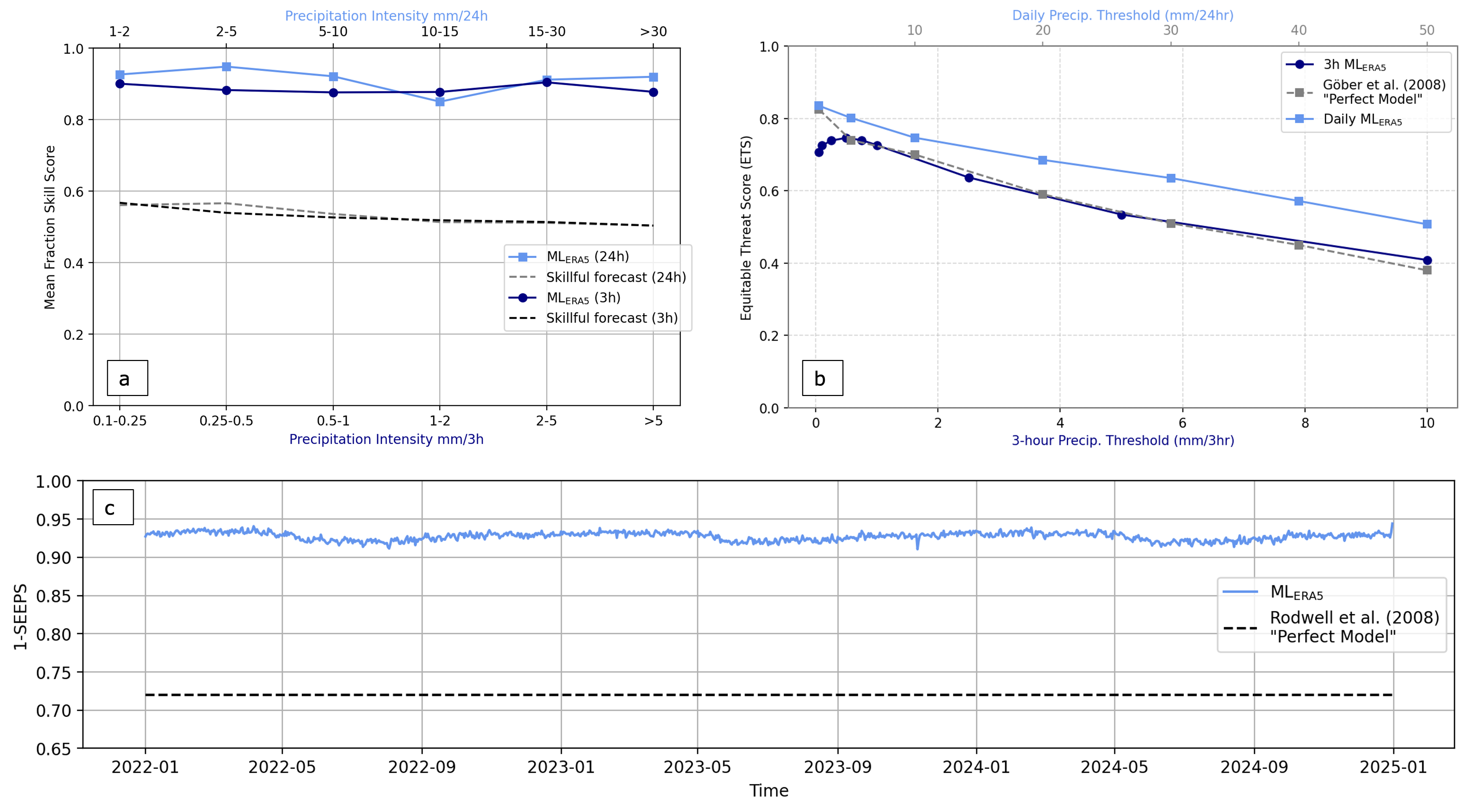}
\end{center}
\caption{Skill metrics for \mle against ERA5 validation over 70N-70S and the test set (2022-2024). (a) Categorical fraction skill score for daily totals from \mle (light blue) and  daily nominal skill thresholds (dashed grey) are plotted for the intensity categories on the top axis. 6-h accumulation categorical FSS from \mle (navy) and nominal skill thresholds (dashed black) are also plotted (bottom axis). (b) Equitable threat score for \mle by intensity threshold for 6-hour (navy, bottom axis) and 24-h totals (light blue, top axis). The dashed grey curve corresponds to the "perfect model" prediction  of \citet{Goeber2008} for daily totals. (c) 1-SEEPS skill score for 24-h totals from \mle (blue) for all times in the test set. The dashed black line represents the approximate 1-SEEPS score of a "perfect model" \citep{Rodwell2010}.}
\label{era5-era5_stats}
\end{figure}

Figure \ref{era5-era5_ex}a and b show 3-h accumulated precipitation examples from \mle (a) and the corresponding ERA5 (b) field beginning at 00 UTC on April 30, 2022. Visually, we see that \mle closely matches the ERA5 field with no obvious blurring. \mle is also able to reproduce many of the fine-scale features present in ERA5 at equal intensity, suggesting it is effectively capturing the ERA5 precipitation distribution. 

Figure \ref{era5-era5_ex}c is a performance diagram comparing the \mle with ERA5 at three intensity thresholds for 3-h and daily precipitation totals average over the globe and the three-year test period. \mle shows an excellent  match with ERA5 at the lightest intensity level for daily accumulations; the performance decreases slightly for medium intensities and 3-h accumulations. There is a modest further loss in performance and a slight bias toward under-forecasting the events for the heaviest intensities (points lying below the 1-to-1 line). The overall performance is quite good considering the POD and FAR are evaluated from stringent point-wise computations.


Figure \ref{era5-era5_ex}d provides a temporal correlation map of 3-h accumulations of \mle with ERA5 over the 2022-2024 test set. The vast majority of the correlations exceed 0.9. The log-transformation of the input data as well as the loss weighting scheme were pivotal for these good results, especially for reproducing high intensity regions.

Figure \ref{era5-era5_stats} gives several globally averaged skill metrics for \mle evaluated against ERA5 precipitation in the test set. Figure \ref{era5-era5_stats}a shows the fraction skill score across six intensity categories for a neighborhood size of $5\times5$ grid cells. Data is not plotted for the most frequent intensity category of dry or almost dry (precipitation less than 1 mm d$^{-1}$). For thresholds below 10 mm d$^{-1}$, FSS values exceed 0.9 and approach a perfect score of 1. At higher thresholds the FSS remains above 0.8, which is well above the nominal skill threshold (dashed black curve) of $0.5+f_0/2$, where $f_0$ is the global mean area coverage of precipitation above each intensity threshold in the verification dataset over a patch of $5\times5$ cells \citep{roberts_scale-selective_2008}.

The equitable threat score (ETS) for \mle 3-h and 24-h accumulations is given in Fig.~\ref{era5-era5_stats}b.  As a reference, scores from \citep{Goeber2008} for ``perfect model" 24-h accumulations in $80\times 80$ km grid cells are also plotted, where the perfect-model score is estimated as that which would be obtained trying to verify European gridpoint observations against the mean of high-spatial-density observations in the same grid cell. The 24-h-accumulated ETS scores from \mle are considerably better than those obtained for the perfect model, implying that if either our, or the ERA5, estimate was compared to actual station observations, it could be difficult to determine which field was superior. Of course since we are comparing gridded predictions from ERA5 with gridded predictions from \mle, a perfect ETS score of 1 is achievable and is the true perfect result.  Nevertheless, from a practical verification standpoint, the performance of both precipitation forecasts would be similar.

The 1-SEEPS score in Fig.~\ref{era5-era5_stats}c gives a measure of overall performance on dry, light, and heavy categories, as determined from the climatology in each grid cell. The \mle scores are approximately 0.94, which are both quite close to a perfect score of 1 and also well above the 0.72 value determined for a ``perfect-model" forecast at 80-km grid spacing \citep{Rodwell2010}. Here the perfect model forecast is again determined as the grid-cell average of actual observations from a dense network and is scored against the station observations in that grid cell.

The results in Figs.~\ref{era5-era5_ex} and \ref{era5-era5_stats} demonstrate that \mle is able to produce precipitation estimates very similar to those generated for the ERA5 reanalysis. We emphasize that these estimates are diagnosed from the 13 fields in Table \ref{tab:variables} without making any attempt to emulate microphysical processes.

\section{Results: Reproducing IMERG Precipitation from ERA5 Inputs}

We now evaluate a similar model (\mli) except that it is trained with IMERG data as the target. Directly training against observational data has the potential to improve precipitation forecasts beyond those obtained via more traditional parameterizations, although we currently lack a perfect dataset against which to train.

We begin with an almost global assessment, omitting only those regions poleward of 70$^{\circ}$ where observations in the IMERG dataset are less complete. Following this, we conduct regional analyses of \mli, focusing on China, where previous studies have shown IMERG to outperform ERA5, and the Eastern United States where Stage IV data provides an independent observational reference. We also assess the diurnal cycle of precipitation, which is known to be difficult for traditional NWP models to capture, both in timing and intensity \citep{hayden2023diurnal, Watters2021DiurnalCycle}. Finally, we assess model skill in capturing extremes and illustrate performance through selected case studies.

\subsection{Global Evaluation}
\begin{figure}[h!]
\begin{center}
    \includegraphics[width=\textwidth]{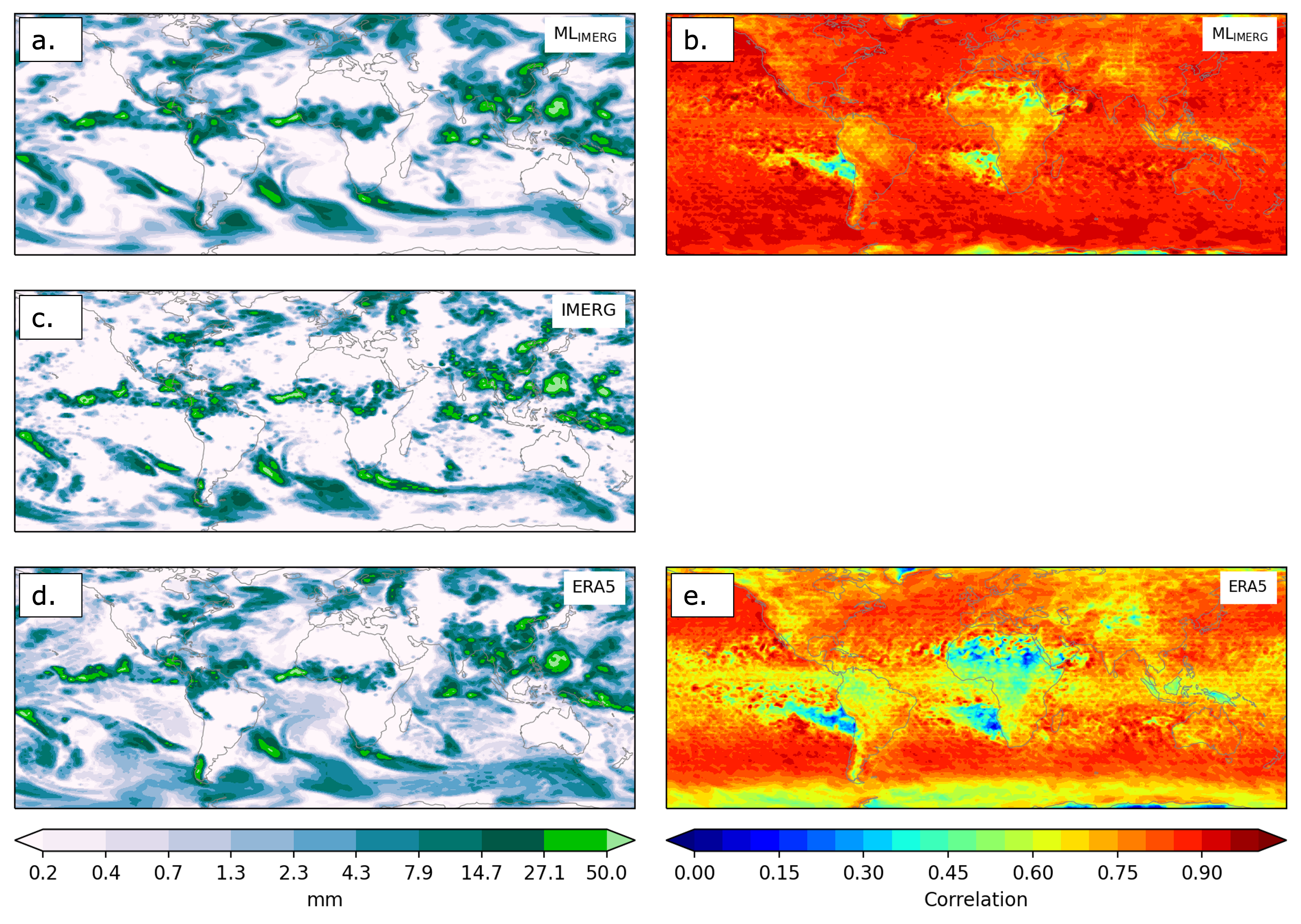}
\end{center}
\caption{Example 24-h totals for July 30, 2023 from (a) \mli, (c) IMERG validation, and (d) ERA5. Temporal
correlations between IMERG and (b) \mli or (e) ERA5. Note that cooler tones show regions of low correlation.}
\label{24hr_example}
\end{figure}

Figures \ref{24hr_example}a, c, and d show a representative example of the 24-h accumulated precipitation field produced by \mli along with the IMERG validation and the ERA5 product. \mli is clearly better than ERA5 at avoiding the overproduction of light precipitation common in non-convection permitting models \citep{stephens_dreary_2010}, although the performance of each method relative to IMERG is harder to quantitatively assess at higher precipitation intensities. The temporal correlation over the three-year test period between 24-h accumulations from each model and the IMERG observations are plotted at every point in Fig.~\ref{24hr_example}b, e. The correlations are much higher for \mli than for ERA5. Since \mli is trained against IMERG we might expect it to do better, but the degree of superiority is perhaps surprising.

\begin{figure}[h!]
\begin{center}
    \includegraphics[width=\textwidth]{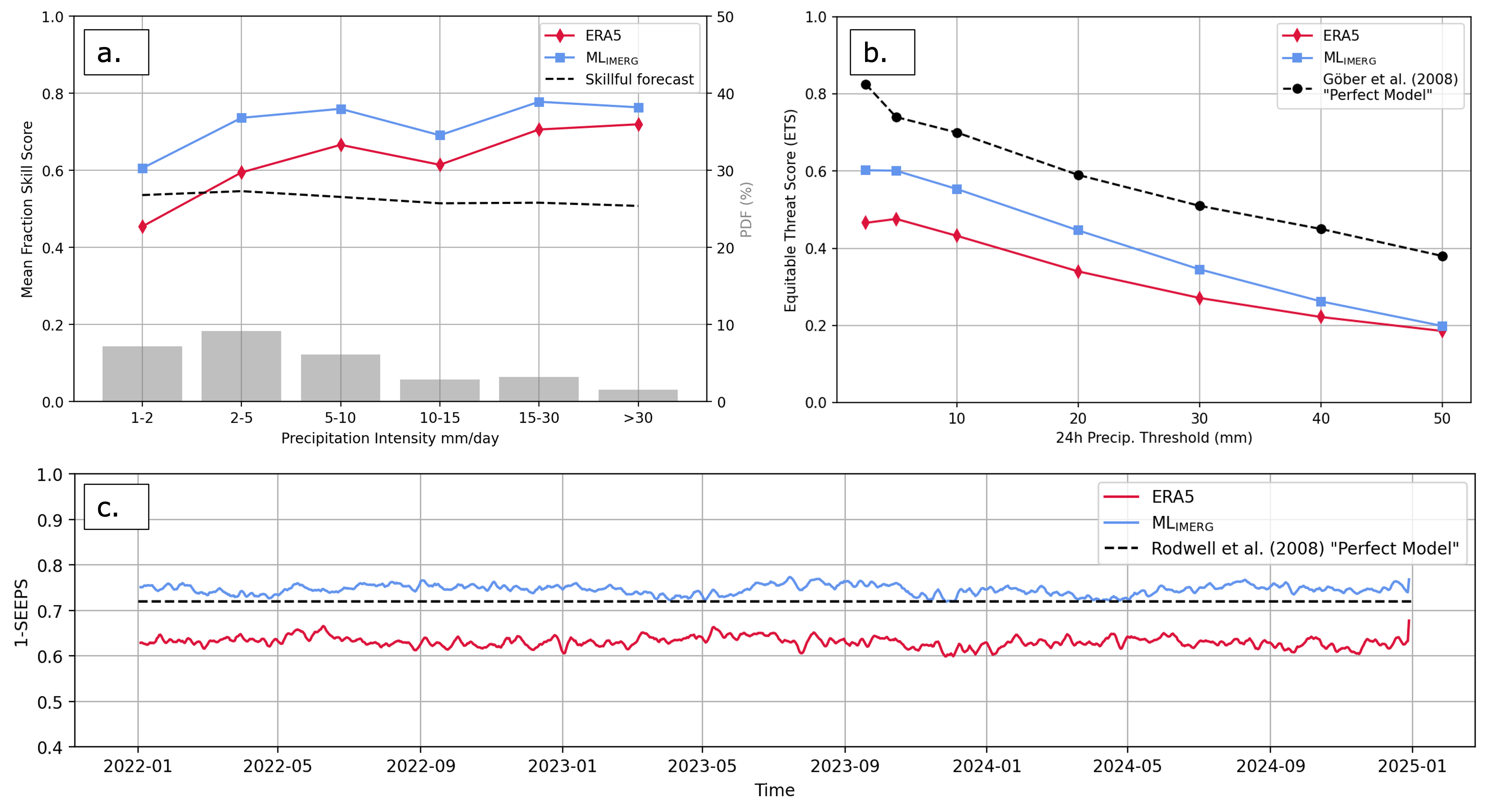}
\end{center}
\caption{As in Fig.~\ref{era5-era5_stats} except \mli (blue) and ERA5 (red) are both scored against the IMERG data: (a) categorical FSS, (b) ETS and (c) 1-SEEPS.}
\label{globe_stat}
\end{figure}

Further quantitative comparison of \mli with ERA5 is presented in Fig.~\ref{globe_stat}, which shows metrics spatially averaged between 70$^{\circ}$ N and 70$^{\circ}$ S over the 2022--2024 test set. Figure \ref{globe_stat}a shows the fraction skill score computed over $5\times5$ grid-cell neighborhoods, binned by threshold intensities.  The FSS diagnosed from \mli is better (higher) than that from ERA5 at all intensity levels.  In contrast to ERA5, \mli scores above the nominal threshold for skillful forecasts at all intensities, whereas ERA5 falls short at the lowest intensity of 1--2 mm d$^{-1}$.

Figure \ref{globe_stat}b compares Equitable Threat Scores (ETS) as a function intensity for 24-h accumulations. \mli demonstrates higher skill than ERA5 at all intensity thresholds, indicating more reliable detection of events beyond random chance. In contrast to the evaluation of 24-h \mle totals against the ERA5 baseline, against the IMERG baseline, both \mli and ERA5 score below the "perfect model" estimates of \citep{Goeber2008}.

Lastly, Fig.~\ref{globe_stat}c shows that the 1-SEEPS score for \mli again exceeds ERA5 in skill, consistently yielding better estimates of dry, light, and heavy precipitation conditions. Note that the 1-SEEPS values for \mli exceed those for the "perfect model" comparison of \citep{Goeber2008}, while the ERA5 results are worse.

\subsection{The Diurnal Cycle}
We now assess the diurnal cycle of precipitation in ERA5, IMERG, and \mli. For increased temporal resolution, the 3-houly totals of \mli, IMERG, and ERA5 were first linearly interpolated to 1-hour resolution and converted to 24-hourly rates (multiplying by 24/3). The data was then mapped to the local solar time at each longitude. Figure \ref{intro_diurnal} plots precipitation rate anomalies as a function of local solar time (LST) over land and ocean for three regions (50N-25N, 25N-25S, and the Maritime Continent). Over land, ERA5 overestimates the diurnal peak intensity of precipitation compared to IMERG in all cases. This peak also occurs too early in the day, preceding IMERG by one to two hours. In contrast, the \mli estimates of peak intensity and timing are much closer to the IMERG observations. Over the ocean, differences in the diurnal signal are less pronounced, and both ERA5 and \mli follow the IMERG cycle reasonably well. This is consistent with the known challenges of representing convective precipitation in NWP models, since the diurnal cycle over land is strongly governed by convection.

\begin{figure}[h!]
\begin{center}
    \includegraphics[width=\textwidth]{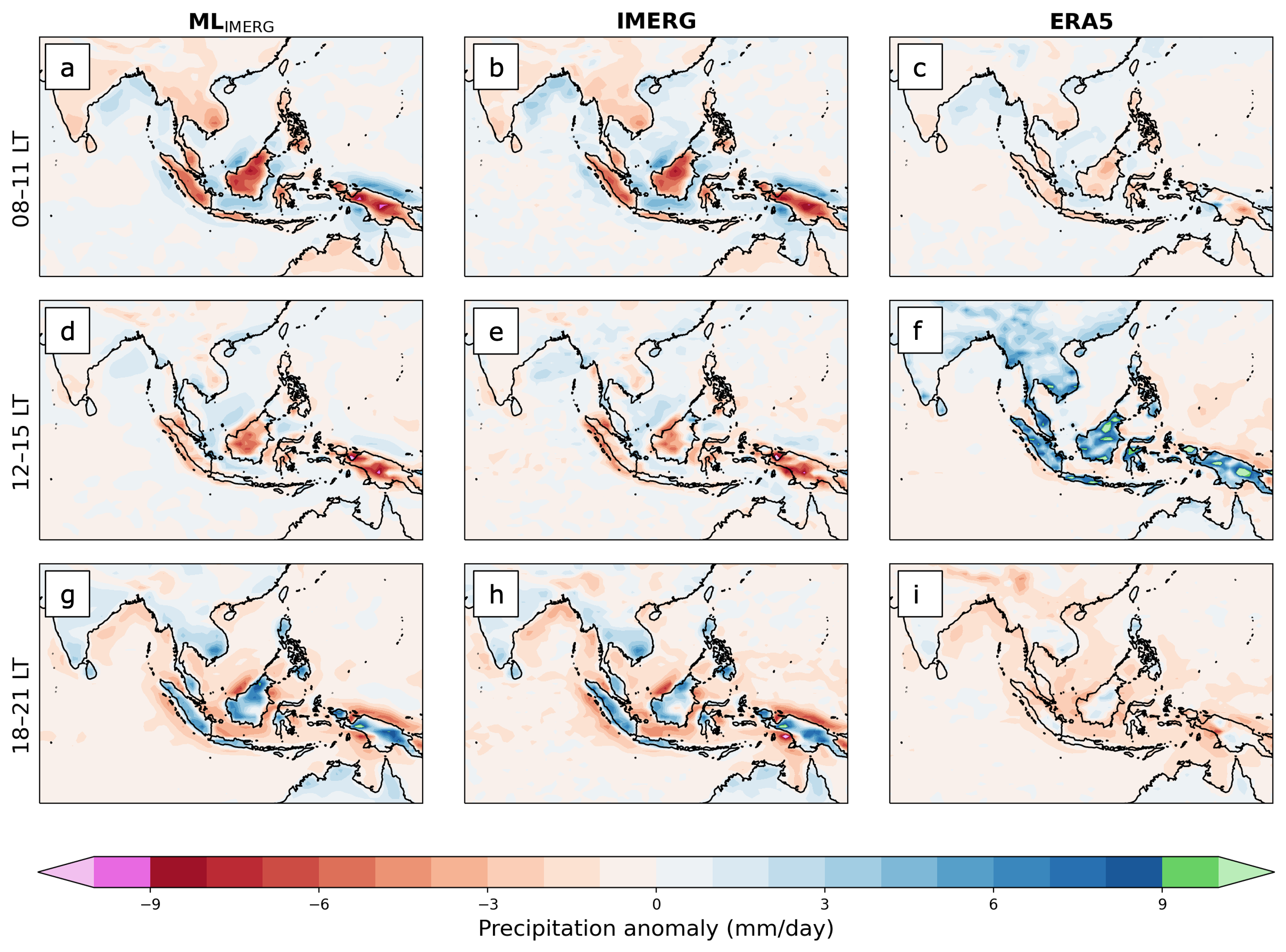}
\end{center}
\caption{Average spatial precipitation rate anomaly for \mli (first column), IMERG (second column), and ERA5 (third column) between local solar times 8-11 (first row), 12-15 (second row), and 18-21 (third row).}
\label{diurnal_maps}
\end{figure}

Figure \ref{diurnal_maps} shows the spatial distribution of 3-hour-averaged precipitation rate anomalies around the Maritime Continent. The first row of Fig.~\ref{diurnal_maps} shows the values  between 8-11 LST, which correspond to the most negative diurnal rate anomaly in the IMERG observations. \mli closely matches the spatial distribution of IMERG precipitation, with a strong contrast between land and ocean around the island coastlines. ERA5 however shows a less pronounced land-ocean gradient, with underestimated positive and negative anomalies. Row two of Fig. \ref{diurnal_maps} gives the spatial distribution of precipitation rate anomalies between 12-15 LST, a period when ERA5 greatly overestimates the anomaly while the agreement between IMERG and \mli is very good. Finally, Fig. \ref{diurnal_maps}g-i shows the precipitation rate anomaly for the time after the peak observed intensity and we again see a similar strong land-ocean contrast in both \mli and IMERG, whereas ERA5 has the wrong sign of anomaly over land. 

\begin{figure}[h!]
\begin{center}
    \includegraphics[width=0.8\textwidth]{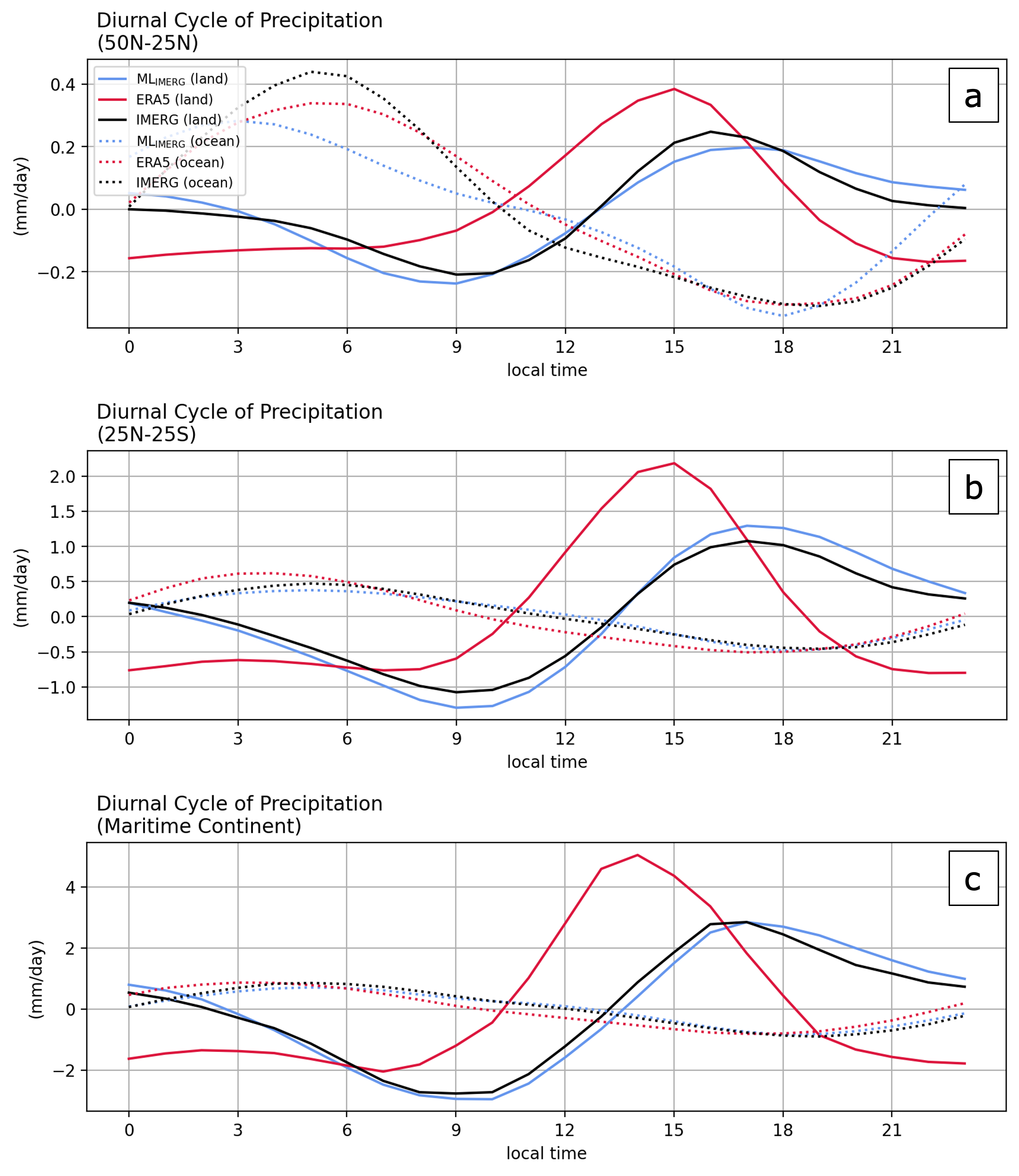}
\end{center}
\caption{Variation of precipitation rate anomalies with local solar time for ERA5 (red), IMERG (black), and \mli (blue). Diurnal cycle is shown for (a) the North mid-latitudes (50N-50S), (b) tropics (25N-25S), and (c) the Maritime Continent over land (solid) and ocean (dotted).}
\label{diurnal_cycle}
\end{figure}

\subsection{Regional Evaluation}
For a more detailed evaluation of model performance, we further assess \mli over two local regions. 
The first is China, where IMERG data has been shown to provide a better estimate of observations at surface stations than ERA5 \citep{jiang_evaluation_2023,wu_statistical_2023,zhou_assessing_2023}, which supports the use of IMERG data for verification in this region.

The second region is the central and eastern United States, where the superiority of IMERG over ERA5 is less obvious than over China. Previous studies indicate that IMERG demonstrates reasonable performance over the United States when evaluated against ground-based observations, particularly during the summer months and across the central and eastern regions where terrain is less complex \citep{Pirmoradian2022}. IMERG has also been shown to represent mesoscale convective systems (MCSs) over the United States with reasonable skill \citep{cui_can_2020}. However, regional and seasonal dependencies remain important: \citet{YousefiSohi2025} found that although IMERG generally exhibits higher skill than ERA5 in warmer, snow- and ice-free regimes, ERA5 tends to outperform the satellite product in colder conditions. For this reason, we evaluate ERA5, IMERG, and \mli against the StageIV radar and rain-gauge dataset \citep{LinMitchell2005} which can serve as both an independent and more accurate verification dataset than IMERG itself.

\subsubsection{Performance over China}
\begin{figure}[h!]
\begin{center}
    \includegraphics[width=\textwidth]{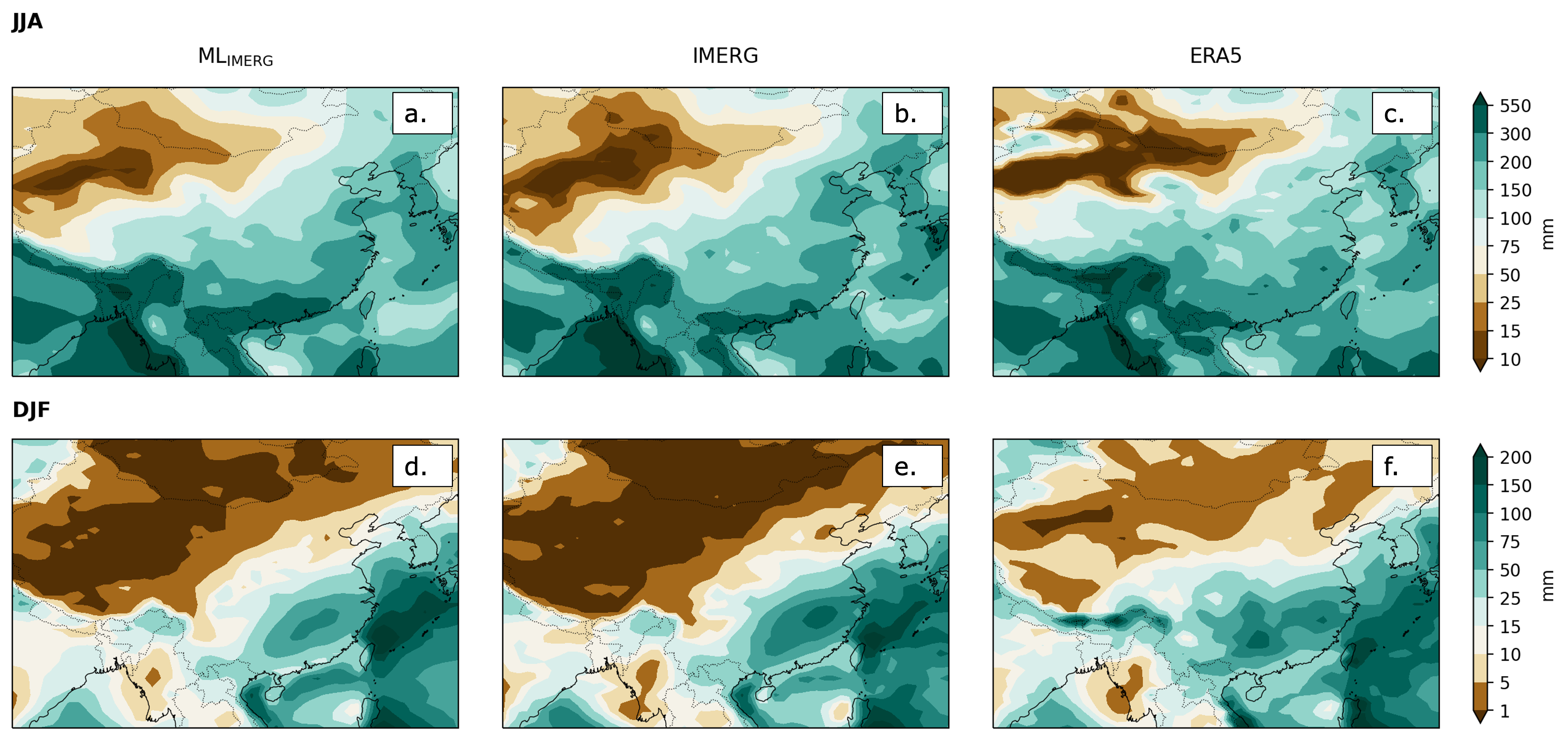}
\end{center}
\caption{(a-c) JJA average monthly precipitation totals over China with \mli on the left, IMERG in the middle, and ERA5 on the right. (d-f) as in a-c but for DJF.}
\label{monthly_china}
\end{figure}

\begin{figure}[h!]
\begin{center}
    \includegraphics[width=0.8\textwidth]{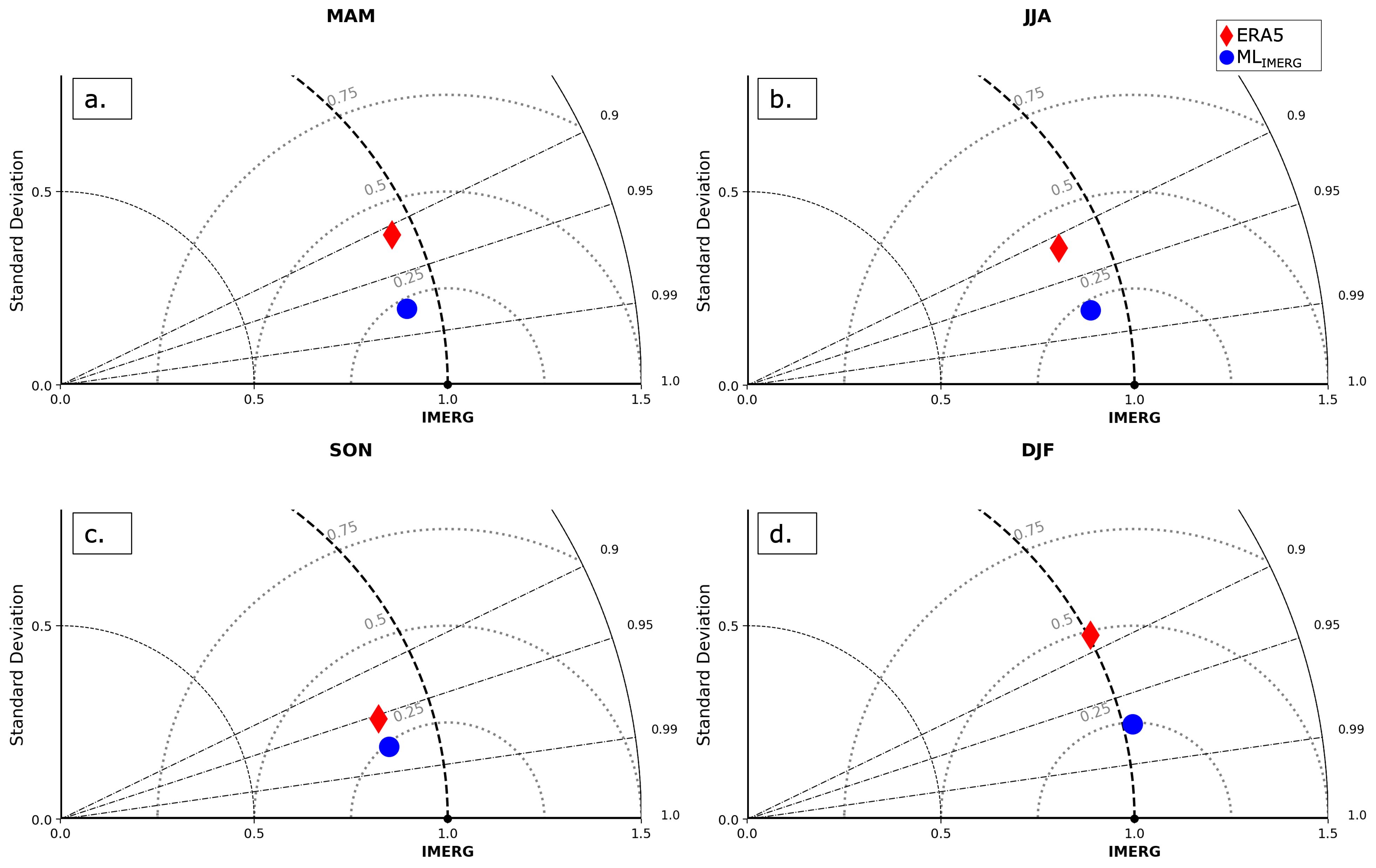}
\end{center}
\caption {Taylor diagrams comparing 3-month totals averaged by season over 2022-2024 from ERA5 (red) and \mli (blue) for the domain shown in Fig.~\ref{monthly_china}. Data are plotted in a cylindrical coordinate system, with radial distance equal to the ratio of the standard deviation of the field under evaluation to the standard deviation of the target field. The angle at which a point is plotted is given by the inverse cosine of the pattern correlation between the field under evaluation and the target. The ``centered RMSE", comparing deviations of each field from its area mean, is given by the distance between a point and the IMERG target point (black diamond) on the abscissa.}
\label{taylor_china}
\end{figure}

Beginning with evaluation over China, the performance of \mli and ERA5 for monthly averaged precipitation during summer (JJA) and winter (DJF) is compared in Fig.~\ref{monthly_china}. 
In JJA, \mli tends to give higher precipitation totals than IMERG, while ERA5 shows lower amounts. As will be quantified in Fig.~\ref{taylor_china}, \mli gives the best agreement. In DJF, the dry region in the northwestern part of the country is better represented in \mli than in ERA5, which overestimates the precipitation.

This comparison is quantified in Fig.~\ref{taylor_china} through Taylor diagrams \citep{taylor2001summarizing} assessing the statistical agreement between IMERG observations and estimates from ERA5 and \mli. \mli consistently shows higher pattern correlation and lower RMSE than the ERA5 estimate. The ratio of the standard deviation over the spatial domain to that in the IMERG data is similarly under-estimated by both \mli and ERA5 in all seasons except DJF, where both estimates are almost perfect.  Overall, Fig.~\ref{taylor_china} suggests the seasonally averaged totals for \mli are significantly closer than the ERA5 estimates to the IMERG observations.

\begin{figure}[h!]
\begin{center}
    \includegraphics[width=\textwidth]{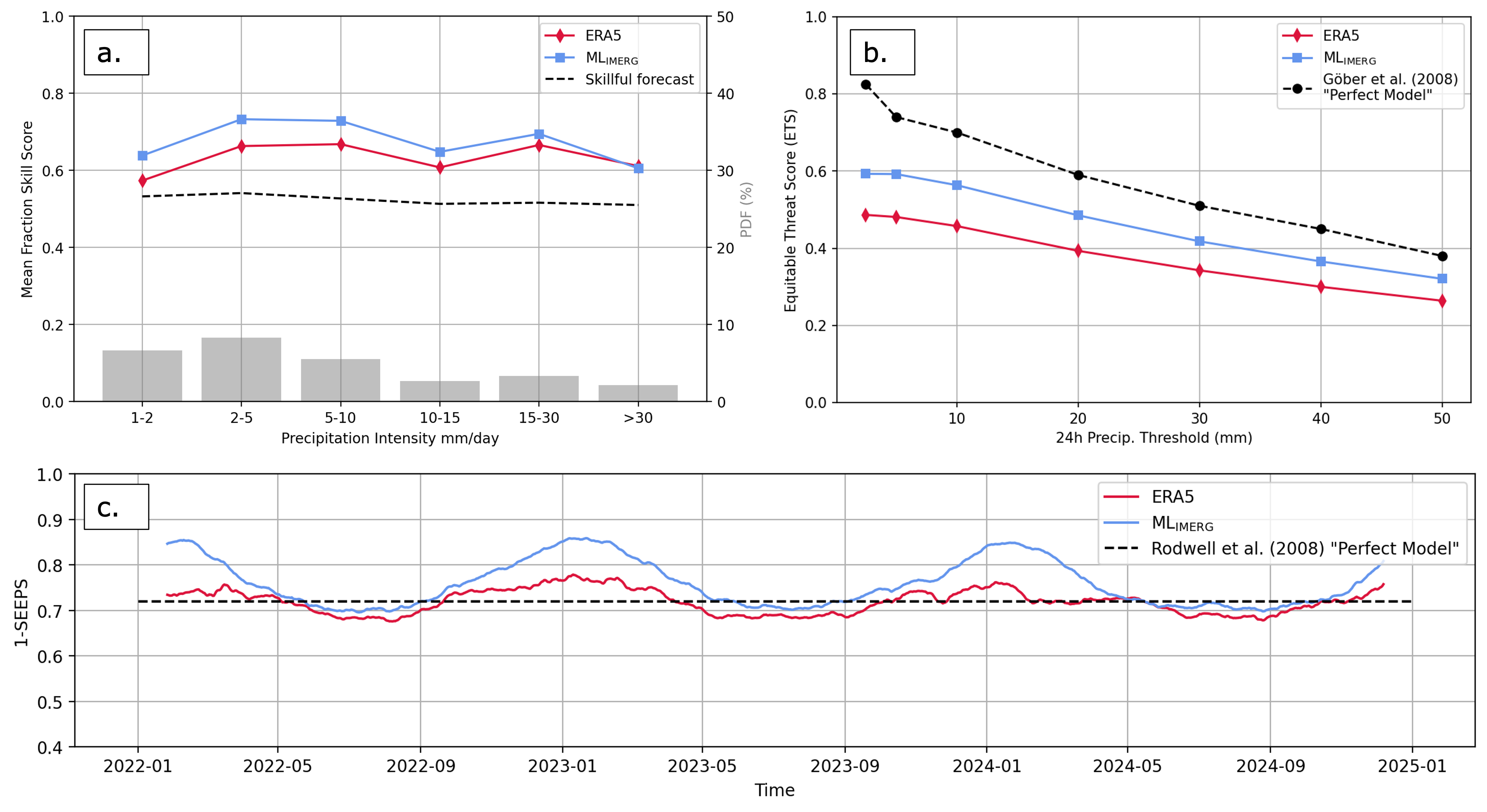}
\end{center}
\caption{As in Fig.~\ref{globe_stat} except domain is that centered on China shown in Fig.~\ref{monthly_china}: (a) categorical FSS, (b) ETS and (c) 1-SEEPS.}
\label{china_stats}
\end{figure}

\mli is superior to ERA5 over China at every intensity category in the fractions skill score as apparent in Fig.~\ref{china_stats}a, although the difference in performance is notably less than that for the nearly global 70$^{\circ}$N to 70$^{\circ}$S region (Fig.~\ref{globe_stat}a). Superiority is also demonstrated for 24-h intensities in the equitable threat score. Finally, comparing 1-SEEPS scores  (Fig.~\ref{china_stats}c), \mli outperforms ERA5 in reproducing the 24-hour totals of IMERG across all days in the test set, with the largest differences in winter. 

\subsubsection{Performance over the United States}

\begin{figure}[h!]
\begin{center}
    \includegraphics[width=\textwidth]{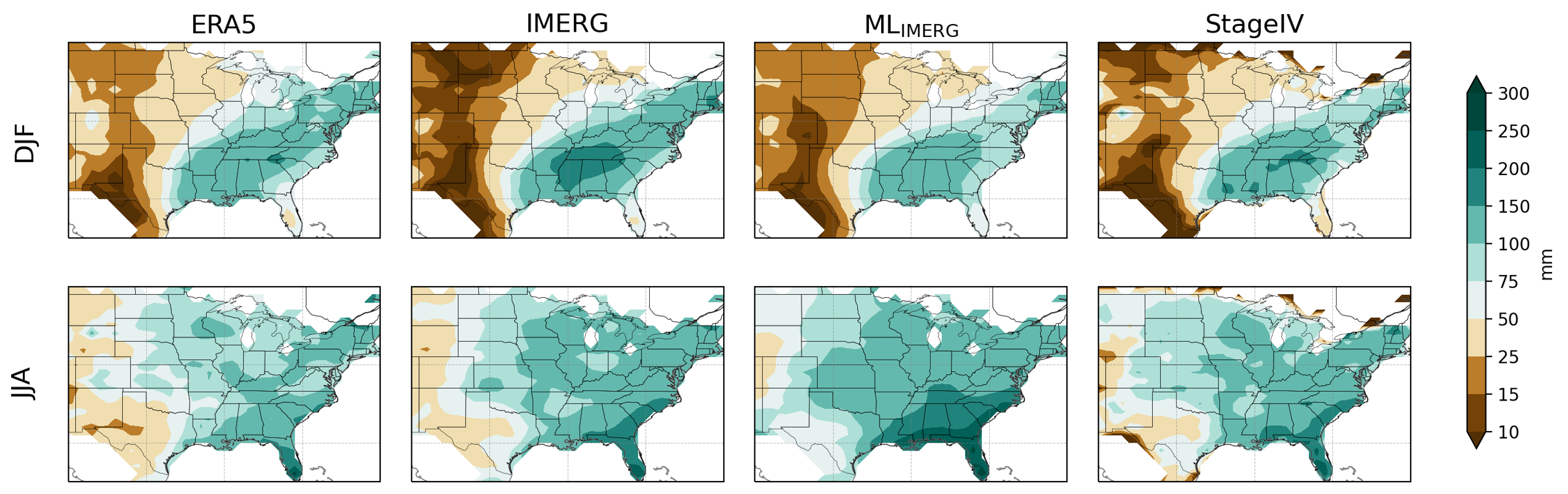}
\end{center}
\caption{(a-h) Average monthly precipitation totals in winter (DJF) and summer (JJA) over the U.S. with ERA5 in the first column, IMERG in the second, \mli in the third, and Stage IV in the last column. The season associated with each row is labeled on the left.}
\label{maps_usa}
\end{figure}

For analysis over the United States, we use Stage IV as the verification dataset. \citet{cui_can_2020} compared IMERG and Stage IV precipitation from MCS over the central and eastern US during the years 2014-2016, concluding they agree reasonably well, notwithstanding some wet bias in the IMERG total precipitation. Figure \ref{maps_usa}a-h gives monthly precipitation totals averaged over each season within the same domain considered by \citet{cui_can_2020} for our test period, 2022-2024. Comparing IMERG (2nd column) with Stage IV (4th column) a modest wet bias is apparent in the central and eastern parts of the domain. In contrast, farther west over the high terrain, a dry bias is clearly apparent in IMERG relative to Stage IV. These differences between IMERG and Stage IV highlight the difficulties in verifying global precipitation fields as well obtaining training data for ML precipitation forecast models.

Turning to the modeled precipitation, the monthly accumulations are too wet in JJA over the southeast, compared to both IMERG and Stage IV. Nevertheless, both \mli and ERA5 provide precipitation estimates in most other areas during both DJF and JJA that differ from Stage IV by roughly the same magnitude as the difference between the IMERG observations and Stage IV. 

\begin{figure}[h!]
\begin{center}
    \includegraphics[width=\textwidth]{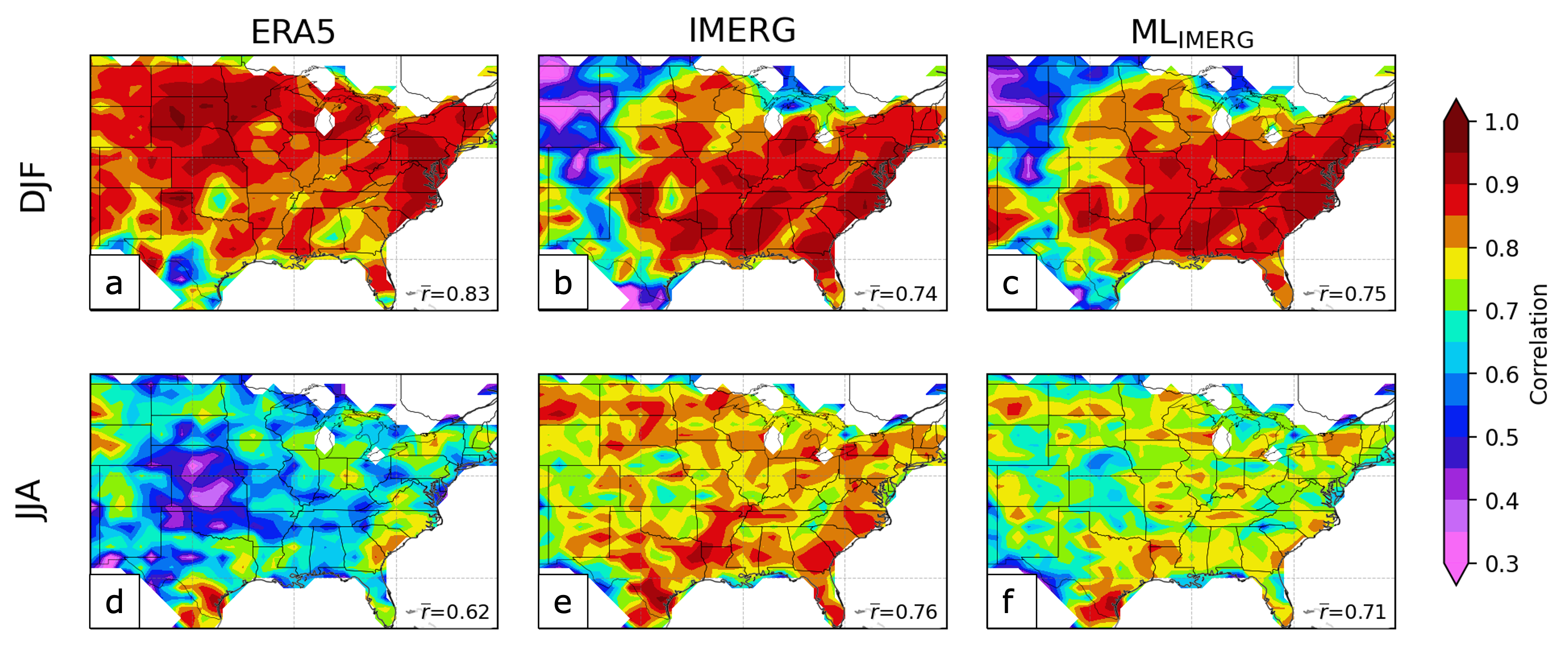}
\end{center}
\caption{(a-f) Temporal
correlations between StageIV daily precipitation totals and ERA5 in the first column, IMERG in the second, and \mli in the third (for the same two seasons as in \ref{maps_usa}).}
\label{corr_usa}
\end{figure}

Temporal correlations of daily precipitation totals with Stage IV for each season are plotted in Fig.~\ref{corr_usa}a-f, with their domain averages noted in the bottom right of each panel. In JJA, when precipitation events are largely dominated by convection, \mli is better correlated with Stage IV observations than ERA mainly in the region of the central plains, with slightly better performance in the eastern half of the country. As also suggested in Fig.~\ref{corr_usa}b and c, the trouble with IMERG and \mli is localized to the higher terrain in the western part of the domain, with \mli performing slightly better than IMERG in this region.

\begin{figure}[h!]
\begin{center}
    \includegraphics[width=0.8\textwidth]{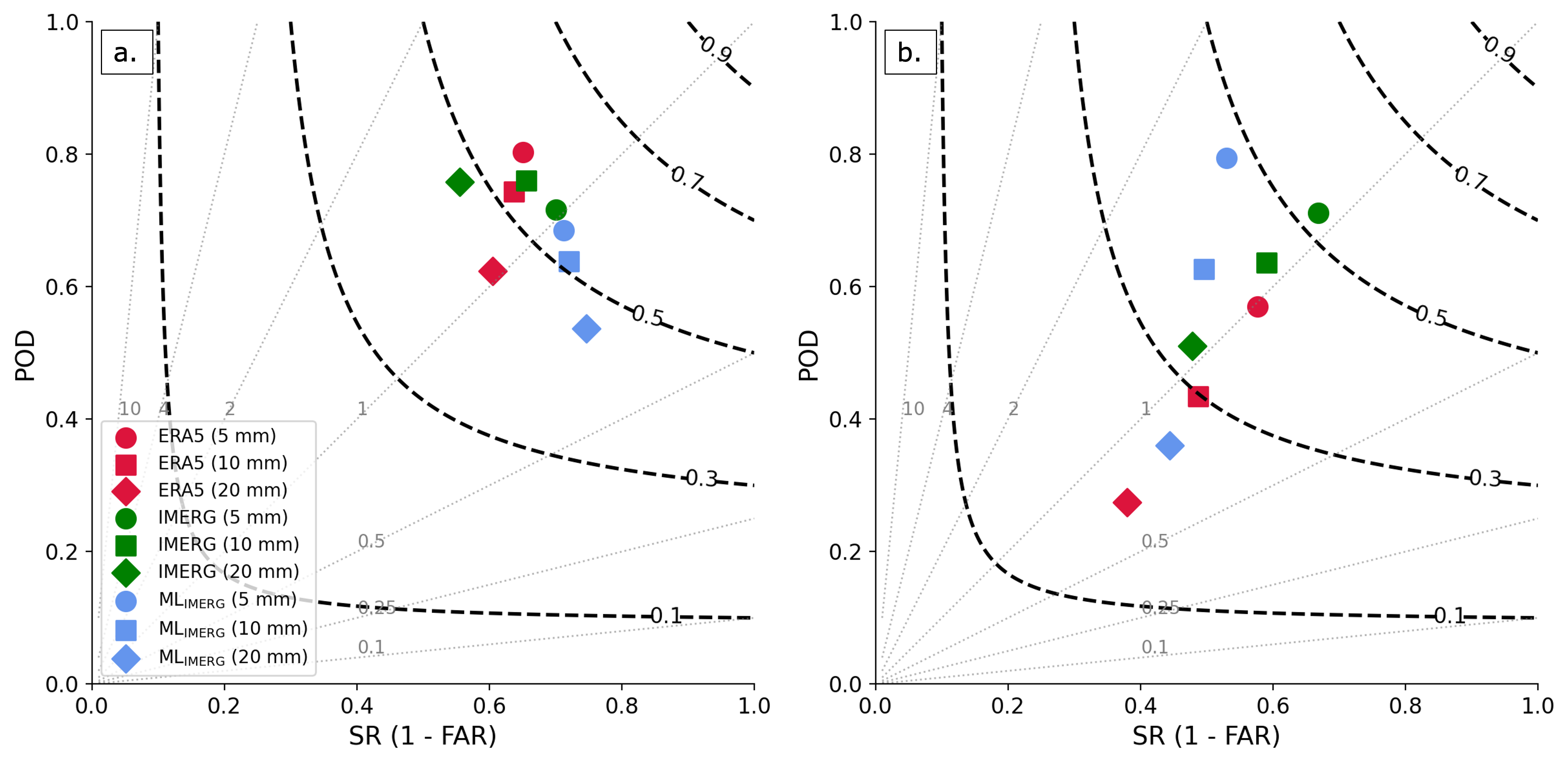}
\end{center}
\caption{Performance diagrams for seasonally averaged 24-h precipitation in (a) DJF and (b) JJA from ERA5 (red), IMERG (green), \mli (blue) validated against Stage IV at three accumulationlevels: 5 mm (circle), 10 mm (square) and 20 mm (diamond).}
\label{US_perf}
\end{figure}

Figure \ref{US_perf} shows performance diagrams for IMERG, ERA5, and \mli validated against Stage IV at three precipitation thresholds. For DJF, the estimates for all sources and thresholds are fairly tightly clustered, with ERA5 slightly better, and \mli slightly worse, than the IMERG values. On the other hand, in JJA, the estimate of light precipitation is much better than that for the heaviest threshold, and IMERG performs better than \mli, which is in turn better than ERA5.

\subsection{Extremes and Case Studies}
\begin{figure}[h!]
\begin{center}
    \includegraphics[width=\textwidth]{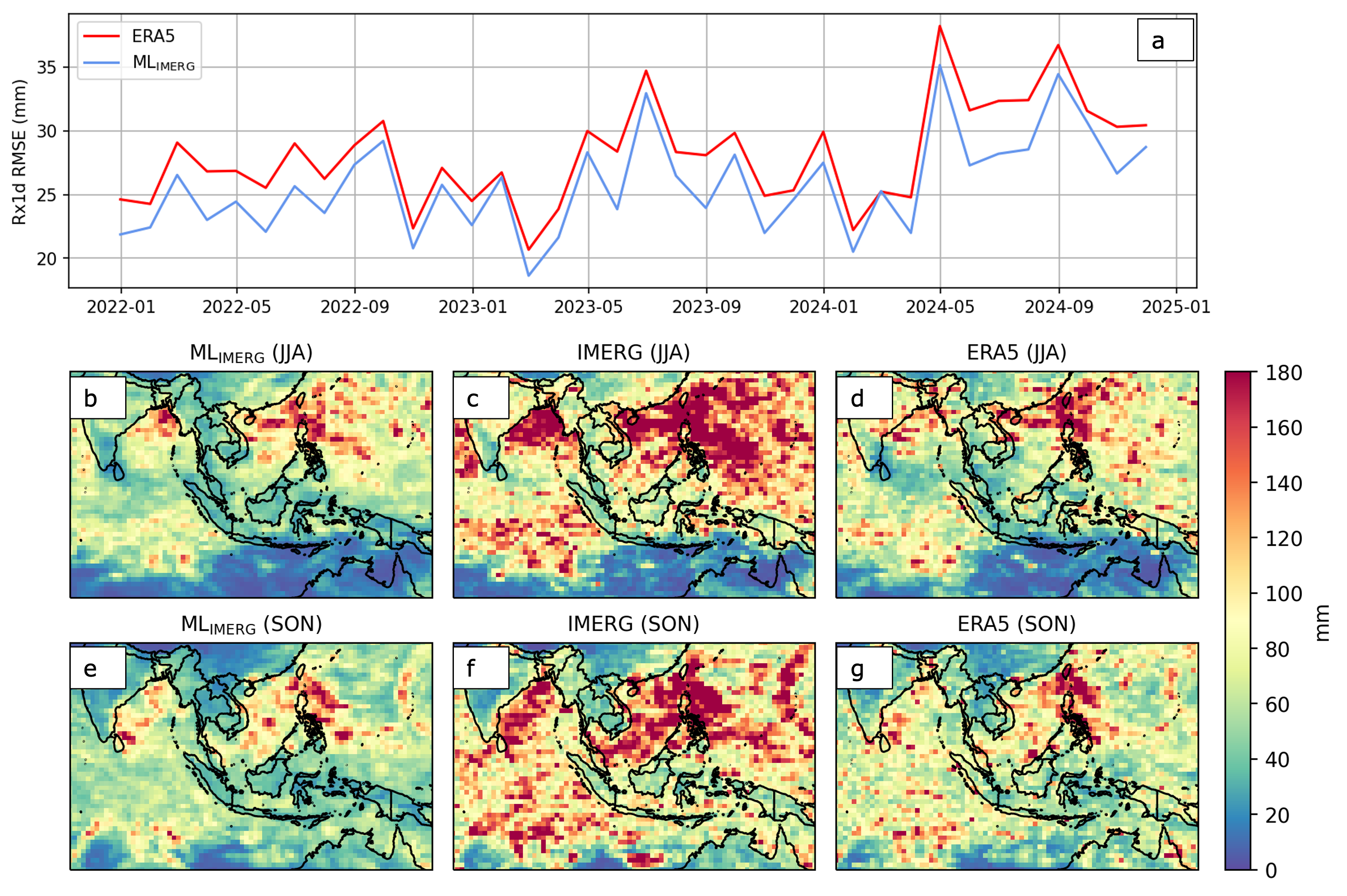}
\end{center}
\caption{(a) RMSE in Rx1d RMSE averaged over the Maritime Continent for each month in the test set. JJA seasonal Rx1d maps for (b) \mli, (c) IMERG, and (d) ERA5 (e)-(f) as in (b)-(d) for SON.}
\label{rx1d}
\end{figure}

As noted previously, a major deficiency of NWP precipitation modeling is the underestimation of extreme precipitation. As a first focus on the climatological distribution of heavy precipitation, consider the maximum one-day accumulation (Rx1d) over monthly and over seasonal three-month periods across the Maritime Continent, which typically has the heaviest precipitation of any similar-sized subdomain on the globe.  Figure \ref{rx1d} compares Rx1d over the 2022--2024 test set from \mli and ERA5 against IMERG.

Using IMERG as verification, the domain-averaged RMSE in Rx1d for every month in the test set is lower for \mli than for ERA5 (Fig.~\ref{rx1d}a).  Examples of the characteristic errors in Rx1d averaged for the JJA or SON seasons are plotted in Fig.~\ref{rx1d}b--g.  Clearly the extent of the most extreme IMERG precipitation is greater than that in either \mli or ERA5.  \mli  captures the intense rainfall over the Bay in Bengal in JJA better than ERA5. In SON, \mli is modestly superior to ERA5 over the South China Sea and along an arc of cyclone-generated precipitation near the eastern edge of the region.

\begin{figure}[h!]
\begin{center}
    \includegraphics[width=\textwidth]{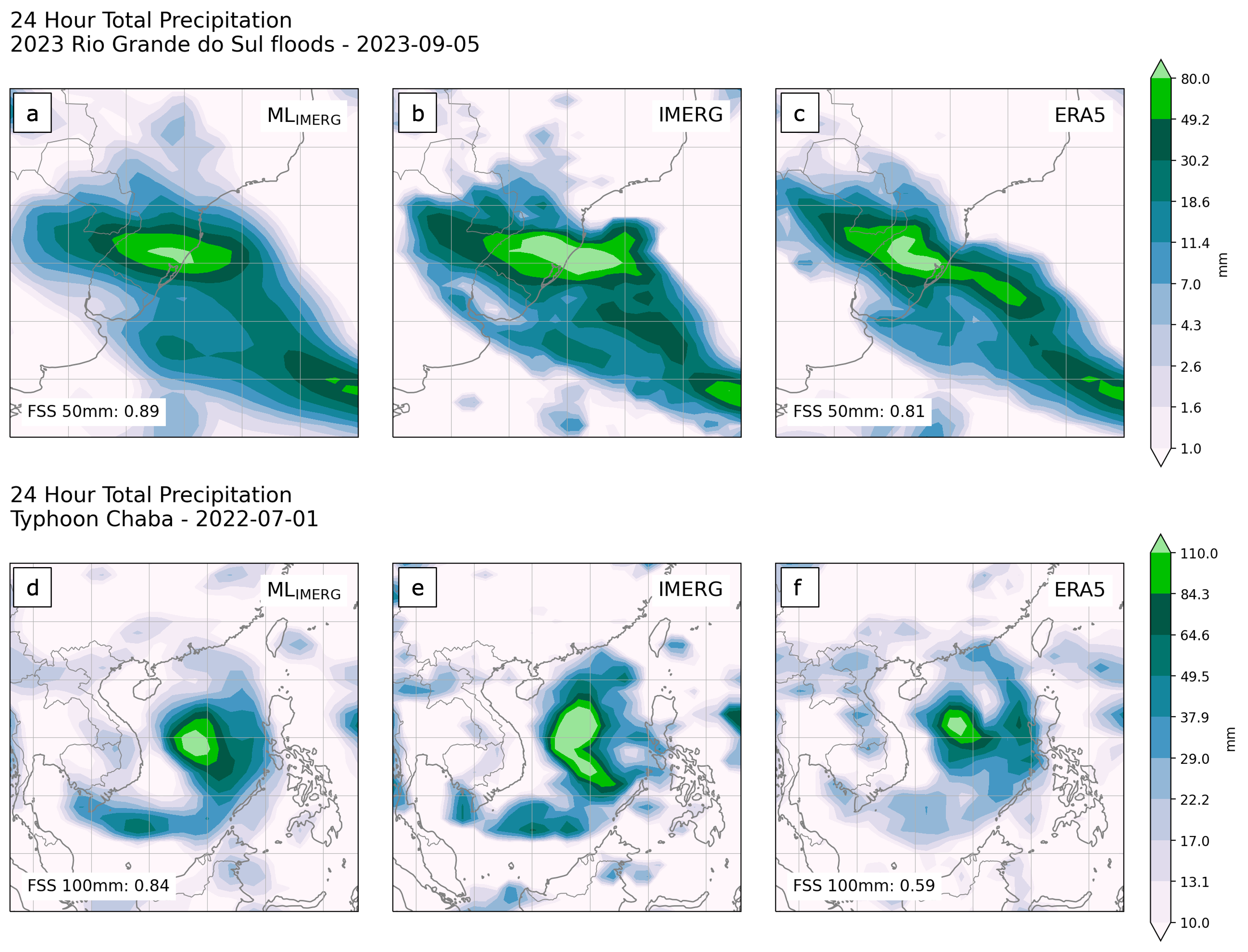}
\end{center}
\caption{Precipitation fields for three different synoptic cases with the \mli estimate on the left, IMERG validation in the center column, and ERA5 in the right. (a)--(c) Brazil flood event 24-h totals for September 5, 2023. (d)-(f) Typhoon Chaba 24-h totals for July 1, 2022. FSS for \mli and ERA5 estimates are also included in the lower left corners of the subplots.}
\label{cases}
\end{figure}

As further representative examples, two heavy precipitation events associated different synoptic-scale systems are compared: a typhoon in the South China Sea and a significant flood event in Brazil. As shown in Fig.~\ref{cases}, in both cases the region of heaviest precipitation is larger in the IMERG data than that from either of the models, with ERA5 slighly out performing \mli in the highest intensity category. The overall precipitation field is, however, better captured by \mli than ERA5 as suggested by visual inspection and quantified by the FSS scores noted in the lower corner of each plot.

\section{Conclusion}

Precipitation in the ERA5 dataset is obtained primarily from short-term model forecasts by ECMWF's IFS as part of the data assimilation process (some radar data from over the US is also assimilated). IFS precipitation is calculated using deterministic parameterizations of microphysical processes and deep convection. Here we have demonstrated that, on the $1^{\circ}\times 1^{\circ}$ spatial scales characteristic of many global climate models, essentially the same precipitation fields can be diagnosed with \mle, a deterministic machine learning model that uses just 13 2D fields from ERA5. Only three of these fields, total-column water vapor, specific humidity at 850 hPa, and outgoing long-wave radiation are directly influenced by atmospheric clouds and moisture. Ablation studies in which we reduced the number of these input fields showed a gradual decrease in performance, suggesting that further improvements in \mle performance could be achieved by inputting more, or perhaps more carefully selected, ERA5 fields.

Because the ERA5 precipitation field is only an estimate of the true precipitation, we explored the potential of our approach to predict estimates of global precipitation from observations rather than trying to further improve the diagnosis of ERA5 precipitation from additional ERA5 input fields. Training our ML model on IMERG data yields \mli, a model whose diagnosed precipitation fields match the IMERG observations at least as well as the ERA5 product across all intensity thresholds, including very light rain and extreme events. While better than ERA5, the \mli fields do not match IMERG as well as the \mle fields matched the ERA5 data, suggesting room for improvement that might be obtained using more input fields or refining the ML architecture.

The IMERG precipitation product has known deficiencies, particularly over mountains and ice-covered surfaces. Unfortunately, there is no universally accepted best global precipitation dataset for training and testing ML models. To assess how \mli precipitation, along with that from ERA5 and IMERG itself, compares to carefully curated regional data, we compared them against NCEP's Stage IV precipitation analysis over the central and eastern US. Overall, IMERG did give the best agreement with Stage IV data, followed by \mli and finally the ERA5 precipitation. Nevertheless, there were periods and locations, such as the high terrain in the west during winter, where the ERA5 precipitation was clearly a closer match to Stage IV values.

Both \mle and \mli compute the total precipitation in much less time than it takes to run CPU-based microphysical parameterization schemes, which pose a computational burden in global atmospheric models \citep{Hong2023_GPU_CAM5_microphysics}. Both \mle and \mli take less than a second per output time to compute the precipitation over the full $1^{\circ}\times 1^{\circ}$ mesh. 

Future work with \mli will explore extending it to the full $0.25^{\circ}\times 0.25^{\circ}$ latitude-longitude resolution of the ERA5 dataset, again with the intention of evaluating the extent to which easy-to-assimilate ERA5 fields can determine the precipitation currently generated by microphysical and convective parameterizations in conventional numerical models. The larger goal in such an effort would be to better understand the extent to which there is potential to improve global climate models without developing ever more detailed parameterizations of cloud microphysics and subgrid-scale deep convection that rely on difficult-to-constrain parameters and difficult-to-observe-and-assimilate moisture variables.

%

%

\clearpage
\acknowledgments

The authors are grateful for discussions with Nathaniel Cresswell-Clay and Zachary Espinosa. This research was supported by the Office of Naval Research under Grants N00014‐22‐1‐2807, and N00014‐24‐12528. This material is also based upon work supported by the NASA FINESST program under Grant No. 80NSSC25K0456. This work was also supported in part by high‐performance computer time and resources from the DoD High Performance Computing Modernization Program as well as a Grant from the NVIDIA Applied Research Accelerator Program. Finally, this work benefited substantially from the barrier‐free high quality ERA5 data set provided by the ECMWF as well as the IMERG data set provided by NASA.


%
%
\datastatement

The training and evaluation data are publicly available through the Climate Copernicus Data Store for ERA5, NASA's EarthData for IMERG, and through NOAA for Stage IV. The source code used for this work will be made available in a publicly accessible GitHub repository upon publication. 

%






%



\bibliographystyle{ametsocV6}
\bibliography{references}

\end{document}